# Stability and Excitations of Spontaneous Vortices in Homogeneous Polariton Condensates

Ting-Wei Chen[1], Szu-Cheng Cheng[2,*], and Wen-Feng Hsieh[1,3,†]

[1]*Institute of Electro-Optical Science and Engineering and Advanced Optoelectronic Technology Center, National Cheng Kung University, Tainan, Taiwan*
[2]*Department of Physics, Chinese Culture University, Taipei, Taiwan*
[3]*Department of Photonics and Institute of Electro-Optical Engineering, National Chiao Tung University, Hsinchu, Taiwan*

Abstract

We study the dynamics of spontaneously formed vortices in homogeneous microcavity-polariton condensates (MPCs). We find that vortices are stable and appear spontaneously without stirring or rotating MPCs. The dip of the vortex core contains some background of reservoir polaritons and the visibility of a vortex is increasing with respect to the pump strength. The vortex radius is inversely proportional to the square root of the condensate density. Excitation energies of vortices at high and low pump powers are finite and zero, respectively. Vortices at low pump powers exhibit the short lifetime.

---

[*] sccheng@faculty.pccu.edu.tw;
[†] wfshieh@mail.nctu.edu.tw

Bose-Einstein condensation is a macroscopic phenomenon from a quantum state due to the coherence of bosons below the transition temperature [1, 2]. The transition temperature of the condensate is inversely proportional to the mass of particles. Excitons coupled to localized light in the microcavity of semiconductor quantum wells form microcavity polaritons whose masses are 5 orders of magnitude lighter than electron [3]. Because of the ultra-light polariton mass the transition temperature of microcavity polaritons could be up to the room temperature [4]. In the past decade scientists took a lot of effort to observe microcavity-polariton condensates (MPCs). Growing interest in MPCs can be attributed to the system being intrinsically out-of-equilibrium determined by the dynamical balance between interactions, pumping and decay [5-7]. The momentum space distributions of MPCs were measured and the accumulation of polaritons in the lowest energy state (or condensation) was observed in planar CdTe-based microcavities [2]. Deng *et al*. [1] measured the second-order coherence function of a MPC and distinguish it from a non-condensate. Due to the continuous pumping and disorders of MPCs, vortices observed in MPCs give the definite evidence of microcavity-polariton condensation [8]. It is surprising that vortices appear spontaneously without stirring or rotating MPCs [6-9]. Non-equilibrium MPCs also show the instability of forming vortices and spontaneous array of vortices without any rotational drive [6-7, 9]. It is important to understand the vortex structure and its stability in studying the dynamics of MPCs. At this moment, there is still lack of a consistent theory to interpret all observed properties of vortices in MPCs.

It is also unique that the vortex radius in MPCs, which is on the order of the healing length, is typically 2 orders of magnitude larger than in atomic condensates and thus is large enough to be observed directly [10]. Spontaneous vortices were observed in incoherently pumped polariton systems in the presence of external confining or disorder potentials [7]. In addition, it can be also generated using a weak external imprinting laser beam in a coherent optical-parametric-oscillator system [9-11]. A minimum power is required for polaritons to acquire enough angular momentum to create a vortex. The vortex lifetime is long/short for high/low excitation powers, respectively [11]. The dip in the vortex core contains some background of polaritons and the observed vortex is not a simple orbital angular momentum state with quantum number $\ell =1$. Due to repulsive interactions of polaritons, the chemical potential of the condensate goes higher and creates a blue shift on the total energy as a high density of polaritons has been injected into the system by raising the pump power [12]. The vortex radius is given by the healing length determined by polariton-polariton interactions and is inversely proportional to the square root of the condensate density or lower polariton blueshift [10, 11]. It was also shown that vortices in incoherently pumped MPCs are stabilized by disorders and have negligible dependence on the excitation conditions [8]. However, in this Letter, we shall show that even in a system without disorder, vortices are stable and arise spontaneously due to the driven-dissipative nature and are sensitive to the excitation conditions.

The vortex state of MPCs is studied through the complex Gross-Pitaevskii equation (cGPE) coupled to the reservoir polaritons at high momenta [5]. This mean-field model for non-equilibrium MPCs is a generic model of considering effects from pumping, dissipation, potential trap, relaxation and interactions. Without considering a potential trap or disorders, we study the dynamics of spontaneously formed vortices in incoherently pumped MPCs. The steady state of the system with a vortex is analyzed under a uniform pump power, $P$. Furthermore, the visibility and core radius of the singly quantized vortex are calculated as a function of pump power. We find that the size of a vortex is inversely proportional to the square root of the condensate density or the pump power above the threshold. Moreover, with increasing pump powers, the core radius and visibility, which are determined by decay rates of the condensate and reservoir polaritons, become smaller and higher, respectively. In addition to studying the steady state of the system with a specific vorticity, the spectrum of elementary excitations around the stationary state is investigated by utilizing the Bogoliubov theory. Because of the non-equilibrium character of MPCs, the excitation frequency $\Omega$ is a complex value [5, 13], whose real part, Re($\Omega$), and imaginary part, Im($\Omega$), represent excitation energy and decay or growth rate of the system. The stability of the singly quantized vortex is justified by Im($\Omega$) provided that Im($\Omega$) < 0. We show that singly quantized vortices can still exist and remain stable under fluctuations even in the absence of stirring, rotating, trapping, or disorders.

In order to study non-equilibrium MPCs, we treat the polaritons at high momenta as a reservoir whose state is determined by its poariton density, $n_R(\mathbf{r},t)$, and employ the cGPE to describe the time evolution and density distribution of the wave function, $\Psi(\mathbf{r},t)$, of the condensate. In the steady state of the system whose chemical potential is $\mu$, $n_R(\mathbf{r},t) = n_R^0$ and $\Psi(\mathbf{r},t) = \Psi_0(\mathbf{r})e^{-i\mu t/\hbar}$. There is no condensate and the reservoir density $n_R^0 = P/\gamma_R$ if $P < P_{th}$. Here $\hbar$ is Planck's constant, $\gamma_R$ is the decay rate of reservoir polaritons and $P_{th}$ is the pump power at the threshold of appearing a condensate. At the threshold, the reservoir density $n_R^{th} = P_{th}/\gamma_R$ is fixed by the balance between the amplification rate R($n_R(\mathbf{r},t)$) and loss rate $\gamma$ of the condensate, i.e., R($n_R^{th}$) = $\gamma$ [5]. When $P > P_{th}$, a condensate appears and the condensate density, defined as $n_c = |\Psi_0(\mathbf{r})|^2$, far away from the vortex core region grows as $n_c = (P_{th}/\gamma)\alpha$, where $\alpha = (P/P_{th}) - 1$ is called the pump coefficient being the relative pump intensity above the threshold. In the mean time, the stationary reservoir density, which is determined by the net gain being zero, is equal to the reservoir density at the threshold pump power, $n_R^0 = n_R^{th}$. Then, the chemical potential of the system is $\mu = gn_c + 2\tilde{g}n_R^0$, where $g$ and $\tilde{g}$ are the strength of polariton-polariton interactions and the coupling constant between the condensate and reservoir, respectively. Throughout this paper we shall take $\tilde{g} = 2g$ under the Hartree-Fock approximation. Given the length unit $\lambda = \sqrt{\hbar^2\gamma\sigma/2mgP_{th}}$ and energy unit $\hbar\omega_0 = \hbar^2/2m\lambda^2$, where $m$ is the polariton mass and $\sigma = 1/[1-(4\gamma/\gamma_R)]$, we can choose the length, time and energy scales in units of $\lambda$, $1/\omega_0$ and $\hbar\omega_0$, respectively. Also rescaling the wave function $\Psi(\mathbf{r},t) \to \sqrt{n_c} \cdot \psi(\boldsymbol{\rho},t)$ and reservoir density $n_R(\mathbf{r},t) \to$ $n_R^{th} \cdot n(\boldsymbol{\rho},t)$, where $\boldsymbol{\rho} = (\rho, \theta)$ is the dimensionless polar coordinate in two dimensions, the cGPE of $\psi(\boldsymbol{\rho},t)$ and the rate equation of $n(\boldsymbol{\rho},t)$ are given as

$$i\frac{\partial \psi}{\partial t} = -\nabla^2 \psi + \frac{i}{2}[\tilde{R}(n) - \tilde{\gamma}]\psi + \alpha\sigma|\psi|^2\psi + (\sigma-1)n\psi, \qquad (1)$$

$$\frac{\partial n}{\partial t} = \tilde{\gamma}_R(\alpha + 1 - n) - \alpha\tilde{R}(n)\frac{\tilde{\gamma}_R}{\tilde{\gamma}}|\psi|^2, \qquad (2)$$

where $\tilde{R}(n) = R(n)/\omega_0$, $\tilde{\gamma} = \gamma/\omega_0$ and $\tilde{\gamma}_R = \gamma_R/\omega_0$.

The steady state of the system under a uniform pump can be obtained by substituting $\psi = \psi_0 e^{-i\tilde{\mu}t}$ and $n(\boldsymbol{\rho},t) = n_0$ into Eqs. (1) and (2), where $\tilde{\mu} = \mu/\hbar\omega_0$ is the dimensionless chemical potential of the system. Using $\tilde{R}(n) = \tilde{\gamma}$ for the stationary condition, we then have the stationary reservoir density $n_0 = \alpha + 1 - \alpha|\psi_0|^2$ from Eq. (2). Therefore, the densities of reservoir polaritons and the condensate are locked together determined by the following time-independent nonlinear Schrödinger equation:

$$-\nabla^2 \psi_0 + \alpha|\psi_0|^2\psi_0 + (\sigma-1)(\alpha+1)\psi_0 = \tilde{\mu}\psi_0. \qquad (3)$$

The chemical potential $\tilde{\mu}$, $\tilde{\mu} = \alpha\sigma + (\sigma-1)$, in Eq. (3) is determined by conditions $\psi_0 \to 1$ and $n_0 \to 1$ in the region far away from the vortex core. Substituting $\tilde{\mu}$ back to Eq. (3), we obtain

$$\nabla^2 \psi_0 + \alpha(1 - |\psi_0|^2)\psi_0 = 0, \qquad (4)$$

which has been used to describe a vortex profile with healing length $\xi = \lambda/\sqrt{\alpha}$ or $\xi = \sqrt{\hbar^2\gamma\sigma/2mgP_{th}\alpha}$ [9, 10]. From $n_c = (P_{th}/\gamma)\alpha$, we find the healing length of vortices is inversely proportional to the square root of the condensate density far away from the core, i.e., $\xi \propto n_c^{-1/2}$. Therefore the vortex radius, $a_V$, is also inversely proportional to the square root of the condensate density which is consistent with the experiments [10, 11].

To find the steady state of a vortex, we just need to solve Eq. (4) and find $\psi_0$ for $\alpha=1$, i.e., $\psi_0^{\alpha=1}(\rho,\theta)$, then the solution $\psi_0^{\alpha}(\rho,\theta)$ for other pump coefficients $\alpha$ can be obtained by rescaling the length scale, i.e., $\psi_0^{\alpha}(\rho,\theta) = \psi_0^{\alpha=1}(\sqrt{\alpha}\rho,\theta)$. We assume that the ansatz of the steady-state solution is $\psi_0(\rho,\theta) = f(\rho)e^{i\ell\theta}$ with $\ell$ representing the winding number of the quantized vortex. Then the total density of the system, $N(\rho)$, is given by $N(\rho) = \chi(\rho) + \kappa(\rho)$, where $\chi(\rho) = n_R^{th}(\gamma_R/\gamma)\alpha f^2(\rho)$ and $\kappa(\rho) = n_R^{th}\left[\alpha+1-\alpha f^2(\rho)\right]$ are the density profiles of condensate and reservoir being determined by the profile $f(\rho)$ of the vortex. We then conclude that the vortex core structure is a combination of the condensate and reservoir polaritons. Due to the reservoir density $\kappa(\rho=0) = n_R^{th}(\alpha+1)$ is not zero at $\rho=0$, the total density of the system is not completely zero at the center of the vortex. Therefore, the dip in the vortex core contains some background of reservoir polaritons, which is consistent with the observed vortex [10]. The visibility of vortex core structure is defined as $V = (N_{max}-N_{min})/(N_{max}+N_{min})$, where $N_{max}$ and $N_{min}$ are maximal and minimal densities of the system, respectively. From $N_{max}=N(\rho\to\infty)$ and $N_{min}=N(\rho=0)$ for a vortex state, the visibility is derived by relaxation rates of the condensate and reservoir polaritons having $V = \alpha(\gamma_R-\gamma)/[\alpha(\gamma_R+\gamma)+2\gamma]$.

By solving Eq. (4) numerically, we can find the profile $f(\rho)$ using the shooting method [14, 15] on the 4$^{th}$ order Runge-Kutta integration with boundary condition $f(\rho)\to 1$ as $\rho\to\infty$. To obtain numerical results we choose physical parameters used in Ref. 8: $m\hbar^{-2} = 1.7\ meV^{-1}\mu m^{-2}$, $\hbar\gamma = 1\ meV$, $\hbar\gamma_R = 5\ meV$, $\hbar g = 0.04\ meV\mu m^{-2}$, $\hbar n_R^{th} = 0.1\ meV\mu m^2$. In Fig. 1, the density profiles of vortex, reservoir, and their summation are shown in unit of $n_R^{th}(\gamma_R/\gamma)$ for $\alpha=6$. The vortex and reservoir densities are inversely related to each other. The condensate density decreases while the density of reservoir polaritons increases as $\rho\to 0$; and the total density at the center of a vortex is given by the density of reservoir polaritons. This is why the observed density inside the vortex core is not absolutely zero in the experiments. Moreover, the vortex radius $a_V$, which is defined from the half width at half maxima of the total density rather than the condensate density, is the actual core radius observed by experiments. For $\alpha = 6$, the core radius $a_V \approx 0.6347\lambda$ is about 5.44 $\mu m$ since the length unit $\lambda \approx 8.5749\ \mu m$, and the visibility of vortex is $V \approx 0.63$. The visibility and vortex radius for other parameters of pump strength are shown in Fig. 2. The visibility and its vortex radius become more clear and smaller, respectively, as the pump power increases. The fitting curve of vortex radii (solid red line) is $a_V = 1.5549\alpha^{-1/2}$ matching with the numerical results (red circles) very well. Therefore, $a_V \propto n_c^{-1/2}$ is confirmed from $n_c = (P_{th}/\gamma)\alpha$ [10, 11]. As we raise the pump power further, the visibility is saturated and approaches the maximal visibility $V_{max} = (\gamma_R - \gamma)/(\gamma_R + \gamma)$ having $V_{max} = 2/3$ for the case of $\gamma/\gamma_R = 0.2$.

After showing the steady-state properties of vortices, we investigate the excitations and stability of the system having a singly quantized vortex with $\ell = 1$. The stationary solutions of the system is perturbed by $\delta\psi = u_q(\rho)\ e^{i(q+\ell)\theta}\ e^{-i\Omega\tau} + v_q^*(\rho)\ e^{-i(q-\ell)\theta}\ e^{i\Omega\tau}$ and $\delta n = w_q(\rho)\ e^{iq\theta}\ e^{-i\Omega\tau} + w_q^*(\rho)\ e^{-iq\theta}\ e^{-i\Omega\tau}$. Substituting the total wave function $\psi = e^{-i\mu\tau}[\psi_0 + \delta\psi]$ and $n = n_0 + \delta n$ into equations (1) and (2) and linearizing them around the steady state, we obtain three coupled Bogoliubov equations [16] that are used to study excitations and the stability of the system whose excited state is labeled by the angular momentum quantum number $q$:

$$-\Delta_+ u_q + (A(\rho) - \mu)u_q + B(\rho)v_q + C(\rho)w_q = \Omega u_q, \qquad (5)$$

$$\Delta_- v_q - (A(\rho) - \mu)v_q - B(\rho)u_q - C^*(\rho)w_q = \Omega v_q, \qquad (6)$$

$$-i\tilde{\gamma}_R[\alpha f(\rho)u_q + \alpha f(\rho)v_q + (1 + \alpha\beta f^2(\rho))w_q] = \Omega w_q, \qquad (7)$$

where, the operators $\Delta_\pm = d^2/d\rho^2 + (1/\rho)d/d\rho - (q\pm\ell)^2/\rho^2$, together with the functions $A(\rho) = 2\alpha\sigma f^2(\rho) + (\sigma-1)[\kappa(\rho)/n_R^{th}]$, $B(\rho) = \alpha\sigma f^2(\rho)$, and $C(\rho) = [(i/2)\beta\tilde{\gamma} + (\sigma-1)]f(\rho)$ are separately defined and the dimensionless coefficient $\beta = R'(n_0)/R(n_0)$ is the change rate of the amplification on the reservoir density. For each pumping scheme, we get excitation frequency $\Omega$ as a function of $q$. There are many excitation states and we will be mostly interested in the branch with the lowest excitation frequency. The decay (Im($\Omega$) < 0) or growth (Im($\Omega$) > 0) behavior of the excitation mode indicates the steady state of the system is stable or unstable, respectively. Note that while there are solutions of three coupled Bogoliubov equations of the form $(u_q, v_q, w_q)$, there should always have solutions of the form $(v_q^*, u_q^*, w_q^*)$ with $\Omega_q \to -\Omega_{-q}^*$.

Discretization transforms the Bogoliubov equations into a matrix equation with eigenfrequencies and corresponding eigenfunctions solved under different pumping schemes, then we find the collective-excitation states and their excitation energies of the system. The low-lying excitation modes with Re($\Omega$) $\geq$ 0 are shown in Fig. 3. The mode patterns are quite different for various pump powers. Note that there exist dispersionless (Re($\Omega$) = 0) and strongly damped (Im($\Omega$) = −5) reservoir mode which is not shown here. This reservoir mode does not affect excitations of vortices much so that it will be neglected in our discussion of vortices. From Fig. 3, except for the cases of lower pump strength close to the threshold, excitation energies for a fixed pump strength increase with the winding number $q$. The excitation energies and decay rates of excitation modes for all $q$'s also become larger as raising the pump strength. We find that finite excitation energies are needed in order to excite vortices in MPCs created by higher pump powers. Once vortices appear spontaneously, they are very stable and not easily destroyed. From the experimental point view, MPCs created by higher pump powers are easier to acquire enough angular momentum to form vortices [11]. There exists a wide range of pump power above the threshold where Im($\Omega$) < 0, implying the singly quantized vortex mode is stable over a wide excitation window. This stability of vortices in MPCs indicates that vortices appear spontaneously without stirring or rotating MPCs that agrees with the experiment [11].

The vortices created by lower pump powers exhibit quite different behavior compared with the vortices created by high pump powers. The low-pump vortices exhibit zero excitation energies and a roton-maxon character for the small positive and negative $q$ values, respectively, as long as $\alpha$ is less than 0.7. The effect of exciting vortices with no excess energy required shows that the low-pump vortices can transform spontaneously into excitation states and show the short lifetime [11]. Therefore, under low pump power, MPCs are difficult to accommodate a vortex and can only be observed momentarily [11, 17].

In conclusion, we develop a theory of spontaneously formed vortices in homogeneous MPCs. We found that densities of reservoir polaritons and the condensate are locked together by the chemical potential. The reservoir density decreases as the condensate density increases and vice versa. Therefore, the dip in the vortex core contains some background of reservoir polaritons, and the visibility of a vortex increases with the pump power. Further increasing the pump power, the visibility would become saturated and approach the maximal visibility. The vortex radius is inversely proportional to the square root of the condensate density above the pumping threshold. Excitations and the stability of a vortex with winding number $\ell = 1$ are also studied. For excitations of all winding numbers, we found a wide stable range of pump power above the threshold having Im($\Omega$) being always less than 0. Vortices with $\ell = 1$ are stable and appear spontaneously without stirring or rotating MPCs. Excitation energies and decay rates of the excitation modes of vortices for all $q$'s become larger as increasing the pump power. The energies of vortices at high pump power are finite and those at low pump power are zero for some $q$ values. Due to the existence of excitations without costing extra energy, vortices at low pump powers exhibit the short lifetime. Our observations in this Letter are crucial and reachable for studying the vortex dynamics of polariton condensates in the future experiment.

We acknowledge the financial support from the National Science Council (NSC) of the Republic of China under Contract No. NSC99-2112-M-034-002-MY3 and NSC99-2112-M-009-009-MY3. S. C. thanks the support of the National Center for Theoretical Sciences of Taiwan during visiting the center.

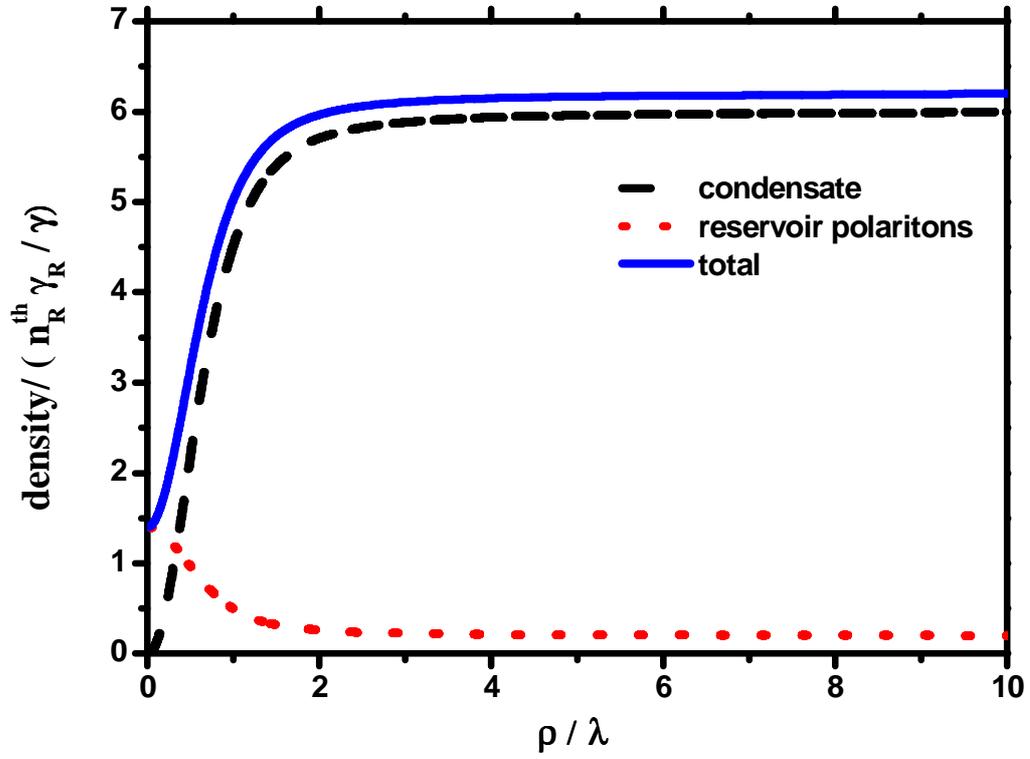

Fig. 1. (Color online) Spatial density profiles. Densities of the condensate and reservoir polaritons are shown by black dashed and red dotted lines, respectively. The blue solid line is the total density of the system.

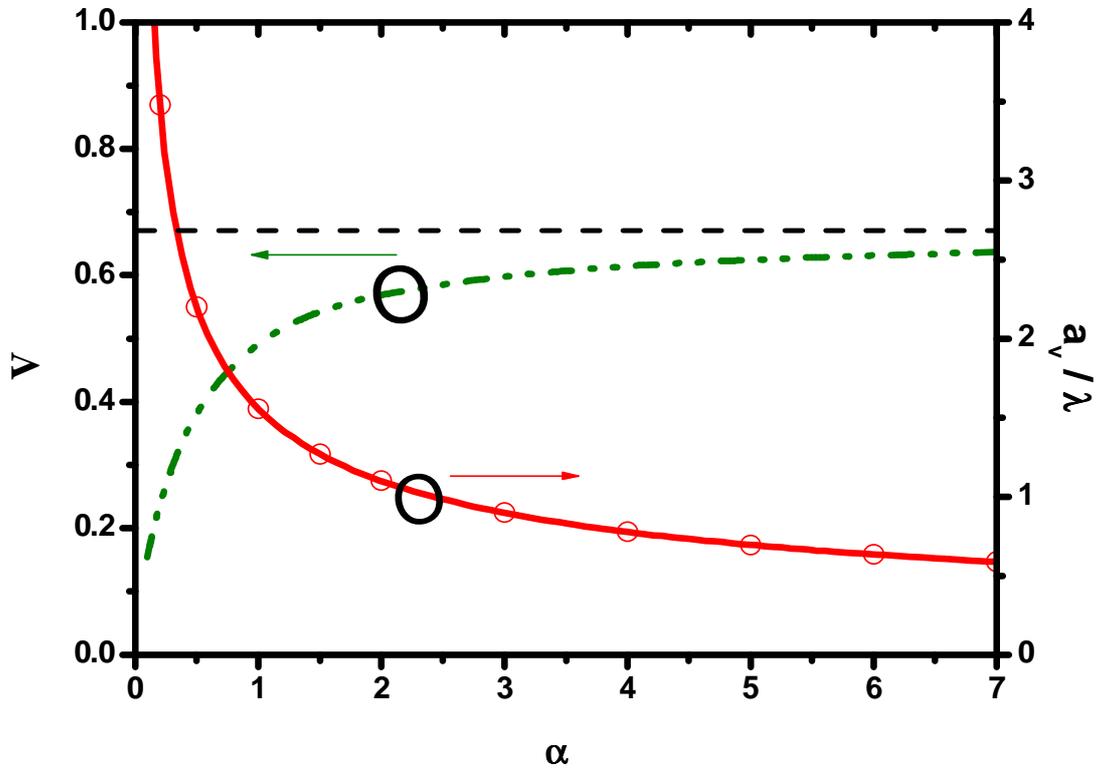

Fig. 2. (Color online) Core radii and visibility of vortices. The core radii (red circles) and visibility (green dash-dotted line) are plotted as a function of pump power above the threshold. Red circles are core radii determined by using the half width at the half maximum of the total density, while the red solid line is the fitting curve of the core radius. The black dashed line indicates the maximal visibility of the vortex.

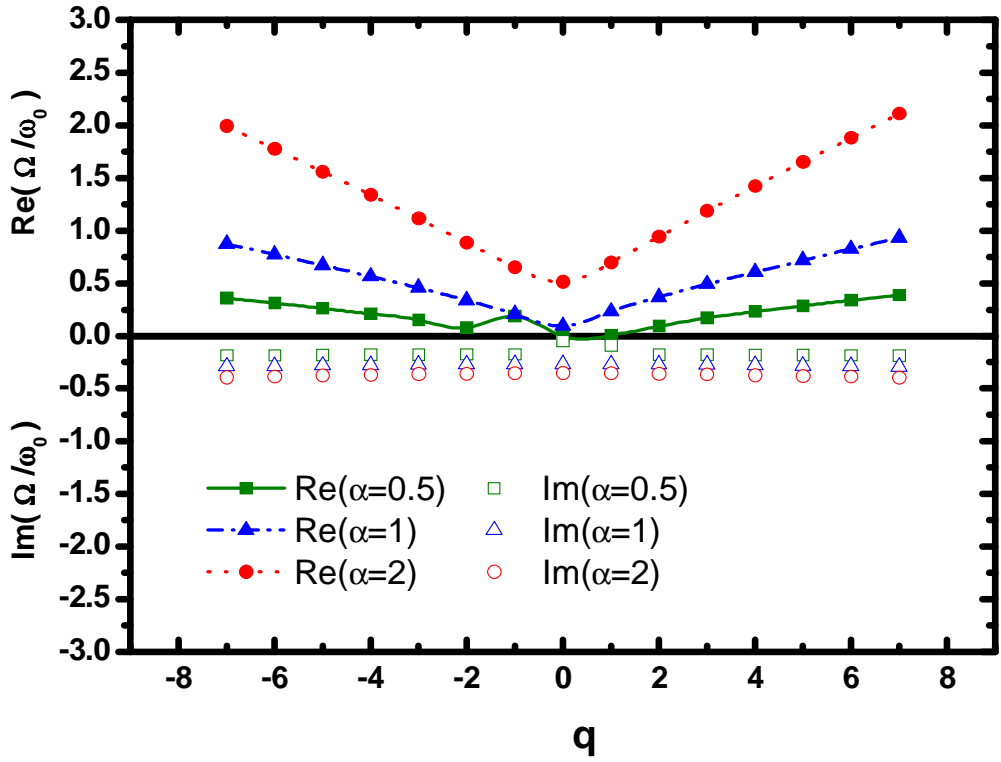

Fig. 3. (Color online) Excitation energies and decay rates of the excitations. Real parts (solid symbols) and imaginary parts (empty symbols) of excitation frequencies of singly quantized vortices are plotted as a function of winding numbers for $\alpha$ = 0.5 (squares), 1 (triangles) and 2 (circles). The change rate, $\beta$, of the amplification on the reservoir density is equal to 1.